# Space and Time

**Sergey V. Shevchenko**[1*] **and Vladimir V. Tokarevsky**[2]

[1] Institute of Physics of NAS of Ukraine

[2] Institute of Chernobyl Problems of Ukraine

## Abstract

In the paper the Space – Time problem is considered as it is seen in the informational conception ("the Information as Absolute conception") comparing with a number of existent physical and philosophical approaches. Since the conception is rigorously grounded (for the conception the existence, the truth, the self-consistence and the completeness of are proven), the notions "Space" and "Time" – which cannot be deduced from physics itself since are Meta-physical notions – obtain reasonable elucidation that is presented in the paper.

## 1. Introduction

This paper presents an approach to one of basic physical problems, namely – to the "Space-Time problem". The approach follows from the informational ("The Information as Absolute" conception) conception (Shevchenko and Tokarevsky, 2007-2013). The conception is rigorously grounded (for the conception the existence, the truth, the self-consistence and the completeness of are proven), and, though it is first of all a philosophical one, it can be applied in the physics directly - since the application reduces in fact to a logical analysis of the problem. By such a way the utmost important in physics notions, "Space" and "Time", - which cannot be deduced from physics itself since are Meta-physical notions – obtain reasonable elucidation presented in the paper below.

---

\* shevch@iop.kiev.ua ; sshev@voliacable.com

To begin with, recall some existent definitions / views of/on the nature of Space and Time (for time durations and spatial distances the words "time" and "space" are used, as a rule, further here).

(Dowden 2005) "... Kant held that Space and Time are the forms of experience, and provide the a priori, intuitive sources of mathematical content… René Descartes had a very different answer to "What is Time?" He argued that a material body has the property of spatial extension but no inherent capacity for temporal endurance, and that God by his continual action sustains (or re-creates) the body at each successive instant. Time is a kind of sustenance or re-creation."

(Markosian 2008) "…Aristotle and others (including, especially, Leibniz) have argued that Time does not exist independently of the events that occur in time. ... The opposing view… has been defended by Plato, Newton, and others. On this view, time is like an empty container into which things and events may be placed; but it is a container that exists independently of what (if anything) is placed in it."

(Dowden 2005) "… Time is what clocks measure. We use our concept of time to place events in sequence one after the other, to compare how long an event lasts, and to tell when an event occurs. Those are three key features of time. Yet despite 2,500 years of investigating time, many issues about it are unresolved. Here is a short list in no particular order of the most important ones: •What time actually *is*; •Whether time exists when nothing is changing; •What kinds of time travel are possible; •Why time has an arrow; •Whether the future and past are as real as the present; •How to analyze the metaphor of time's flow; •Whether future time will be infinite; •Whether there was time before the Big Bang event; •Whether tensed or tenseless concepts are semantically basic; •What the proper formalism or logic is for capturing the special role that time plays in reasoning; •What neural mechanisms account for our experience of time; •Why time is one-dimensional and not two-dimensional and •Whether there is a timeless substratum from which time emerges." [Note, that some of the listed problems can be directly addressed the Space notion.]

(Peterson and Silberstein 2009): "This problem stems from two competing notions of time. The first, originally suggested by Heraclitus, …is the view that only the present is real; both the past and the future are unreal… However, with the advent of relativity, a different stance, whose primary ancient proponent was Parmenides of Elea … was translated into the language of relativity of Hermann Minkowski in 1908 to suggest that time and space should be united in a single, four-dimensional manifold. Thus arose the notion of a 4D "block universe" … in which the past, present, and future are all equally real."

(McTaggart 1908) "…It would, I suppose, be universally admitted that time involves change…" [But, after considering temporal ["A", "B", "C"] "series of events] he concludes "…We cannot explain what is meant by past, present and future." … "Our ground for rejecting time, it may be said, is that time cannot be explained without assuming time"

(Britanica Online Encyclopedia: Space) "… Space**,** a boundless, three-dimensional extent in which objects and events occur and have relative position and direction." [When it is possible that] "… Space is not an independently existing thing but merely a mathematical representation of the infinity of different spatial relations that particles may have to each other. In the opposing view,… Space is an independently existing thing, and what facts about the universe there may be do not necessarily coincide with what can in principle be established by measurement".

(Newton 1686) "…Absolute, true and mathematical time, of itself, and from its own nature flows equably without regard to anything external, and by another name is called duration: relative, apparent and common time, is some sensible and external (whether accurate or unequable) measure of duration by the means of motion, which is commonly used instead of true time"

"…Absolute space, in its own nature, without regard to anything external, remains always similar and immovable. Relative space is some movable dimension or measure of the absolute spaces; which our senses determine by its position to bodies: and which is vulgarly taken for immovable space ... Absolute motion is the translation of a body from one absolute place into another: and relative motion, the translation from one relative place into another"

And, at last:

(Minkowski 1952, 75-91): "…The views of space and time which I wish to lay before you have sprung from the soil of experimental physics, and therein lies their strength. They are radical. Henceforth space by itself, and time by itself, are doomed to fade away into mere shadows, and only a kind of union of the two will preserve an independent reality" …"We should then have in the world no longer *space*, but an infinite number of spaces, analogously as there are in three-dimensional space an infinite number of planes. Three-dimensional geometry becomes a chapter in four-dimensional physics. Now you know why I said at the outset that space and time are to fade away into shadows, and only a world in itself will subsist".

Einstein (1961): "…There is no such thing as an empty space, i.e., a space without field. Space–time does not claim existence on its own, but only as a structural property of the field... It appears... more natural to think of physical reality as a four-dimensional existence, instead of, as hitherto, the *evolution* of a three-dimensional existence".

(Westman, Sonego 2009) "…According to general relativity, the concept of space detached from any physical content does not exist."

(Rovelli 2003): "…. Change is not described as evolution of physical variables as a function of a preferred independent observable time variable. …To put it pictorially: with general relativity we have understood that the Newtonian "big clock" ticking away the "true universal time" is not there."

(Butterfield 2002): "…Physicists are able to compactly summarize the workings of the universe in terms of physical laws that play out in time. But this convenient fact should not trick us into thinking that time is a fundamental part of the world's furniture … But it, too, … is a convenient fiction that no more exists fundamentally in the natural world than money does."

This citing doesn't, of course, comprise all multitudes of nuances of all existent philosophical/ physical schools – that seems impossible. Nonetheless it, in general, outlines the mainstreams in physics and philosophy. As well as we don't consider here advantages and disadvantages of the views above in detail, and note only some straightforward consequences. First of all – the views/ approaches aren't distinct in details only, the differences are fundamental. Space and Time are a priori products of the [human's] consciousness; are objective things that are absolute and independent on anything in Matter; are objective but don't exist without change [in itself or of positions] of material objects; at least Time (time) is unreal at all – since a " logical analysis of time series" leads to contradictions; or since any physical situation is "described in terms of a functional relation among equal footing variables [which aren't some derivatives of time]" (Rovelli 2008). In the relativity theories Space and Time are objective, but depend on each other ("only a kind of union of the two will preserve an independent reality"), constituting the pseudo-Euclidian manifold ("Minkowski spacetime") in special relativity theory (SRT); and, besides that, they are objective, but exist only in presence of Matter (in GRT).

We don't consider further such proofs of unreality of Time as, e.g., "logical contradiction" of the notions "past", "present", and "future" since (if the time is continuous) the present is impossible inasmuch as it is simultaneously the past and the future. That is, in fact, some remake of Zeno's aporias in an application to the time notion, which, in turn, have became be resolved, in certain sense, with creation of quantum mechanics. In the paper we discuss how the Space and Time notions follow from the informational conception in comparison with the views above and try to answer on a number of the questions from the "Dowden's list". For convenience we give a brief introduction into the conception.

## 2. Space and Time in the informational conception

To continue with, make a couple of epistemological remarks: (i) - any indeed new knowledge can be obtained by a human only by empirical way – A. Einstein (cited by A. Salam (Salam 1979)): "Pure logical thinking cannot yield us any knowledge of the empirical world; all knowledge of reality starts from experience and ends in it." (Wittgenstein 1921, point 6.1262): "Proof in logic is merely a mechanical expedient to facilitate the recognition of tautologies in complicated cases."

And (ii) – practically any implications from any experiment - e.g., "Nature laws" - in reality always remain be as some hypotheses only – since are based on the necessary but insufficient criterion of the reiteration of given experimental outcomes in given experimental conditions. Logically a scientist can only *believe* in that the next outcome will be "in accordance with the theory".

But there is *the unique exclusion* - in the case when somebody experimentally observes a few of things that are inherent to the information: a set and a number of logical rules that define the concatenations between the set's elements.

### 2.1. The informational conception - a brief introducing

So from an experiment resulting in "discovering" of an information one can rather simply deduce a number of important inferences, which, in contrast to any other empirical inference, *aren't beliefs* and, at that, *are always true*.

1. Any information [always] exists as a unity of logical concatenations that are applied to elements of some set and of some descriptions [including some rules that control interactions between] of the set's elements. Or, in other words, any information exists as some unity of Logos and "inert" (fixed) information since to concatenate the elements is necessary to define them and the rules.

2. Null (empty) set for any specific set contains this set totally since to define the null set is necessary to point out that this set doesn't contain every element of the specific set.

3. So the global empty set (GES) contains full information about everything that exists, doesn't exist, can exist and cannot exist – i.e. the GES entirely contains all absolutely infinite information, which, in turn, is contained in fundamental absolutely infinite Set "Information".  Besides, the GES differs from any specific empty set. A specific empty set "is placed" always outside its generic set, for example, the specific empty set for the set "Earth + Moon" – semantically "there is neither Earth nor Moon" is placed in any point of space (and everywhere in the Set else) outside Earth and Moon. But the GES – semantically at first sight can be expressed as "there is no anything" - is an information and so belongs (is an element of the Set) to the Set; in contrast to any specific set the Information Set is closed relating to Her empty set. However the GES semantic form above isn't totally correct since, when there is no anything else, there is the information that there is no anything. So correct the GES's expression is the infinite [dynamic] cyclic statement "there is no anything, besides the information that there is no anything, besides…" It is rather probable, that this semantic distinction (any specific empty set   is a fixed information) of the GES from the specific empty set allows the GES to remain inside its host Set.

4. Any information cannot be non-existent – that follows from existence of the GES, which, by definition, exists (is true) when there is no anything else.

5. To define any element from the Set is necessary to point out all distinctions of this element from every element of Set – i.e., any element of the Set contains all Set entirely. And, eventually,

6. From the points 1-5 above follow  existence, truth, self-consistence and  completeness  of the informational conception (more – see Shevchenko and Tokarevsky 2010).

Though – since we fundamentally cannot go out the conception – we cannot prove its uniqueness. That, in principle, can create some difficulties to apply the conception in the practice. But the premise about existence of something outside the Set seems now as unessential – insomuch it is clear that anything, what is outside the Set, has no any informational presentation. Thus, since material objects evidently can be characterized by some information, and since the Information Set is a unique object that doesn't require anything else besides itself for Her existence, – we have, as it seems, quite of sufficient reason to suggest, e.g., that Matter in our Universe (as well as Universe itself) is (are) some [infinitesimal comparing to the Set, even they are infinite] sub-set(s) of the Set.

At least in first approximation Universe comprises three subsets – Matter, Alive and Consciousness, where two of the lasts contain also possible alive and consciousness beings outside Earth (including some beings that possible live in other Consciousness's subsets outside the human's one). This paper deals with Matter only, so we point out the main property of Matter that defines it in Universe: *Matter is an informational [sub-] set where every concatenation of its elements, i.e., - every interaction of some material objects, proceeds necessarily as an exchange by true information exclusively*.

Besides we should make some additional remark. From the text above follows, that the Set contains all information about anything "in any time of its existence/ evolution" always, since the Set cannot be non-existent. Absolutely everything in the Set "has happened absolutely long time ago"; at that "it is happening always". Including, e.g., that the scenario of entire evolution of our Universe "was known, has happened, and is happening always" also, – so the "past", the "present" and the "future" states of Universe always existed, exist and will exist simultaneously. So the notion "Time" isn't applicable in whole Set *in certain sense*. But for some separated subsets in the Set that isn't so, and, for example, anybody observes in Her subset "Matter" given local events "in present time"; it turns out to be that for an observer (as well as for any other material object in Matter, though) there exists a cause that "realizes" only one given "virtual" scenario as some objective (not necessarily observable) material process of Matter's evolution. We don't know – what is this cause, but further consider here the Matter's evolution as when at given time only one "Matter's state" subsists (sometimes can be observed/ measured) - as that follows from common human's experience. The problems, which arise when quantum mechanical effects become be essential, seem as non-surprising, if one recall on the very paradoxical properties of the Information Set; however, they require additional study and are outside scope of this paper.

*2.2. Space and Time*

From the human's experience follows that material objects always (and "only truly") interact in accordance with a number of "Nature laws". So Matter is an ordered dynamical system, i.e. is something like a computer. This premise isn't new, of course - it is enough to recall, e.g., Pythagoras's "All from number" and Plato's "All from triangles" doctrines, first strings of Bible Genesis, etc.; more specific hypotheses that our Universe is a large computer appeared practically at once with the appearance of usual computers (more Refs see, e.g., Shevchenko and Tokarevsky 2012). But that were only some hypotheses, which have not necessary grounds (besides, of course, the case when one considers a Creation of Universe as of a logical structure from nothing by some omnipotent judicious Being, Who "made the computer and established a program code").

Except, though, C. F. von Weizsäcker's 1950-54 years idea of the quantum theory as of a theory of binary alternatives ("UR- theory"). Weizsäcker "…Mathematically, … had just stumbled about the fact that *SU(2),* the quantum theoretical symmetry group of a binary alternative, is locally isomorphic to the three-dimensional rotation group *SO(3)* in Euclidean space" (Lyre 2003). That was indeed an objective hint on inherent binary structure of Matter what necessitates also the 3- dimensional structure of space. But it doesn't answer on the question – nonetheless, what is Space at all? And – analogously – what is Time?

In informational conception the answers are straightforward – Space and Time are some logical structures that realize themselves in Matter as a rules/ conditions/ possibilities governing (together with some other rules/ conditions) "the executing of the program code(s) in computer "Matter"; first of all, they single out different informational structures (IS) in Matter – particles, bodies, Galaxies, etc. At that Space singles out/ controls, first of all, fixed information. Time controls the dynamical changes/ evolution of the ISs; *defining* that in any cause-effect chain of changes the cause always happens before the effect (cf., e.g., Reichenbach 1928). This rule sometimes seems as not too convincible. For example, Dowden (Dowden 2005) cites an objection to defining of time with causal order – "there is nothing metaphysically deep about causes preceding their effects; it is just a matter of convention that we use the terms "cause" and "effect" to distinguish the earlier and later members of a pair of events which are related by constant conjunction". In the informational conception such an objection is incorrect – the Time rule isn't "a matter of convention", it realizes itself as a *logical [semantic] order*, when logical cause-effect concatenations are inherent base of any informational structure. Moreover, the rules are universal for any separated by some way subset in the Set totally – including outside given Universe: in any cause-effect event there is some "time interval" between the cause and the effect or "space interval" between some informational patterns. These intervals can be infinitesimal, but *never are equal to zero exactly*. At that they can be different for different events (elements of a system) and so are some parameters characterizing the events (systems).

Note, besides, that as some *outer* (and *primary*) to Matter rules/ conditions/ possibilities, Space and Time are "absolute", are independent on each other and exist "forever", including – before [possible] Beginning and after [possible] End of Matter's evolution in our Universe.

*2.2.1. Spacetime*

As some possibilities Space and Time are realized in Matter as 4-dimesional "spacetime". The space in spacetime, in accordance with von Weizsäcker's idea, is 3- dimensional to form some logical structures that are based on a binary logic. The *time in spacetime* ("coordinate time", "τ- time") is one-dimensional, all (space and time) directions are mutually independent and so should be mutually orthogonal, the spacetime should be *Euclidian*. Thus a rather plausible informational model in physics (Shevchenko and Tokarevsky 2007, 2012) can be put forward. In the model material objects are some [closed for rigid bodies, including - for elementary particles] algorithms that run on "hardware" of some analogs of Weizsäcker's "Urs", i.e, - on "fundamental logical elements" (FLE). FLEs constitute a dense FLE – lattice in the spacetime – some analogs of Penrose's "spin-network units"(Penrose 1971), "causal set" (Sorkin 91), "Space-time points in causal space" (Finkelstein 69), etc. An FLE has at least four degrees of freedom, i.e. can "flip" in 3 spatial directions and 1 time direction. A sequence of FLE flips is the motion of "flipping point" – or of flipping FLE - in spacetime; when FLEs, after an impact with transmission to some FLE non-zero momentum, are flipping also in an specific [closed] algorithmic order, at that a specific particle occurs. Since after any impact FLEs' flipping never stop, when the flipping rate is stable and identical for all FLEs, all existent particles/ bodies move *always* in 4-D spacetime *having identical speeds* (the speed of light).

Here we again should make an additional remark. In the Set Her elements are, of course, singled out also. Since any informational connections between any elements can be reduced to sequential choosing of alternatives in a graph, the Set - though being *absolutely infinite* - is in certain sense "discreet". So anything in Matter, including space and time, is/are discreet also. The information in the Set can propagate with an infinite [but not with "absolute infinite", i.e., there is no simultaneity] speed; an example: if, e.g., an electron hits into a Galaxy that is on, say, 10 billions light years apart from Earth, in every space point outside the Galaxy, including on Earth, immediately "virtual" (but true) information appears: "an electron have hit into given Galaxy". Corresponding time interval for a cause-effect chain "an event in Galaxy – the change of information in a point" is infinitesimal. However, propagation speed of real information in Matter – as that follows from human's experience - is limited. Why that is so? There is no answer till now – that is a physical problem for a further study. But now it seems rather plausible to suggest that in Matter in our Universe there is some minimal time interval – as well as – some minimal space interval; and these intervals are FLE flip time, $\tau_P$, and the size of the FLE, $l_P$. If we suggest also, that a change of a state of FLE ("0" or "1") on 1 bit is analogous to the change of half-integer spin of a material particle, i.e. it is seen as the change of angular momentum or [physical] action on the elementary quantum of the action, i.e. – on [reduced] Planck constant, $\hbar$, then we obtain 3, rather plausible fundamental, constants that define the evolution of FLEs – as well as of everything in Matter; namely: Planck time, $\tau_P$, Planck length, $l_P$, and reduced Planck constant.

Since the basic in Matter are some closed algorithms, there is the fundamental difference between Space and Time, which "works" outside, in certain sense, spacetime. Independent material bodies can be placed in space in an arbitrary order and so can move in any direction, including – in direct and reversal one. For Time that isn't so. On one hand, when it is possible to move only in time, any motion in any other direction logically is also motion in time, since any motion is a cause-effect process; so corresponding "true time intervals" have only one direction by definition (are positive by a convention). But to realize a closed logical sequence, e.g., when creating a particle, is necessary to have some logical rule/possibility that allows implementing reverse logical orders. Such rule/possibility differs from the *true ti*me *rule*, but, nonetheless, in certain sense they are similar since logical dynamic sequences. In Matter the rule is realized as "*coordinate ti*me" ("τ- time") above.

For example, photons move in space only (have infinite time dilation in standard formulation) and so don't move in τ- time; but since space-motion is a cause-effect process, they must move in some other time also - i.e. in true time. Besides, all other [material] objects move in space and in τ- time practically in any direction with speed of light and so move (with, of course, speed of light) in the true (or "absolute") time always also. So the rule "Time" has at least two application in Matter – "coordinate time", τ, and "absolute time", *t*. "Coordinate time", defining order of events, is similar to absolute tome, but in a number of traits it is like as the space (e.g., FLE really flip with equal to the space footing in τ-time's direction) also, constituting with the space Matter's Euclidian spacetime, when "absolute time" *isn't a coordinate* of the spacetime and controls any – space or τ-time (or space and τ-time) motion of a particle. At that, of course, both times (as the spacetime as a whole, though) are "absolute", i.e. don't depend on any – being at rest in space or moving – material object or on any "reference frame" (cf. "Euclidian relativity" - Montanus, 1999, Nawrot, 2007 and a number of other authors).

Space and Time [as universal rules] and space and time [as the realizations of the rules and as the parameters of separation in space and in time material objects and their motion] are independent on anything, but their application to a moving FLE give rise to some interdependence of the FLE's flips. Any FLE can flip (and so the particle moves) in spacetime in any direction – including exclusively in time ($\tau$), or exclusively in [arbitrary] space (*X*) direction – always having one speed, which is – experimentally - equal to speed of light. But if flipping FLE moves in some ($\tau$,*X*) spacetime direction, $\tau$ and *X* components of its speed – and corresponding $\tau$, *X* components of FLE flipping rate - become be lesser then speed of light (lesser then $1/\tau_P$), since the FLE's flips' rate cannot be greater then inverse Planck time's value. Thus, insomuch material objects are some algorithms constituting from always flipping FLEs (more, again – see Shevchenko and Tokarevsky 2012), those algorithms become run slower, e.g., in $\tau$ direction if an object moves also in space. In existent theories that is known as "the time dilation" that tenably reveals itself when some material object moves in space with speed be near speed of light.

*2.2.2. Spacetime in existent theories*

Historically, the electrodynamics and mechanics of fast moving objects were developed in 1887 -1905 years mainly by Voigt, FitzGerald and Lorentz (further – "VFL -theory"), when the equations that concatenate spatial and temporal variables were obtained ("Lorentz transformations" (LT)). Though some interdependence of the variables seemed as rather evident, the authors, who didn't doubt in truth of Newton's view on absolute spacetime, considered such a situation as some mathematical trick ("a convention" by Poincare) that relates by some way to some local dynamical properties of Matter's bodies.

In 1905 -1908 years Einstein – Minkowski special relativity theory was developed. In contrast to VFL- theory, the SRT have attributed the LT to some kinematical *spacetime* nature – when "space by itself, and time by itself, are doomed to fade away into mere shadows". Thus, in fact, two additional (to well known two Einstein's 1905 ones) postulates were included implicitly into the theory: (i) – the SRT is a global theory, i.e. some *variables* in the LT became be the coordinates of spacetime *x, y, z, t*; which are defined for space and time on infinite intervals, and (ii) – there is no absolute reference frame(s) in spacetime, all inertial reference frames are equivalent. The basic parameter of the SRT is so called "invariant interval", *s*, and to keep the interval be invariant any [inertial] frame transformation must be a rotation of frame's axes – there can exist simultaneously a number of the frames, where *time-axes have different directions*, what is impossible in Newton spacetime. That immediately has led to self-contradictories in theory, first of all – to the "twin paradox". This paradox is known mainly as the "clock paradox", but it has another variant - as the "energy paradox". "Twin- traveler" at motion not only becomes be younger then the "homebody one", besides, if he measures homebody's energy, then he obtains that this energy is rather large – when the homebody spent no fuel. Further, from positing that all reference frames moving in Matter with different speeds are totally equivalent directly follows that Matter has corresponding different real masses – when seems as rather evident that unique in our Universe Matter must have unique mass, even this mass is infinite, etc.

This "paradox" isn't a paradox; in reality it reveals the fact that the SRT is self-inconsistent and so it cannot be "resolved" in framework of the SRT. Thus any existent "the paradox resolution" always contains some logical or physical flaw, which can be found at some analysis. As, e.g., seems a standard one (Wiki), where the paradox becomes "be resolved" since

"…In special relativity there is no concept of *absolute present*... ...The notion of simultaneity depends on the frame of reference, so switching between frames requires an adjustment in the definition of the present. If one imagines a present as a (three-dimensional) simultaneity plane in Minkowski space, then switching frames results in changing the inclination of the plane." (see Fig. 1)

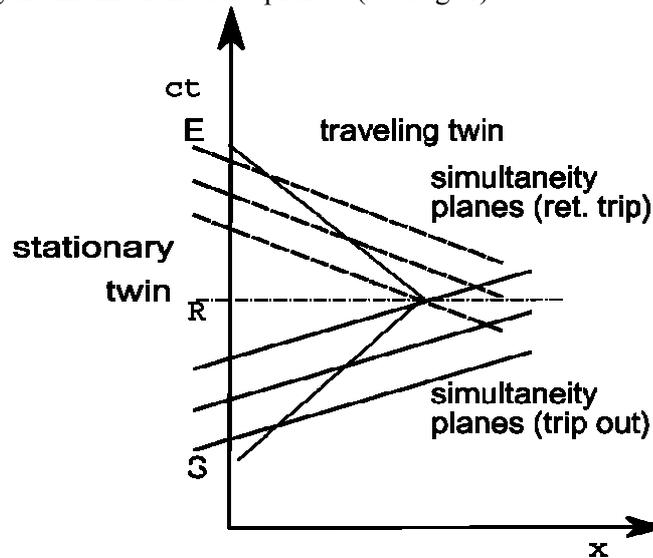

Fig 1. (Wiki) Minkowski diagram of the twin paradox. Time is relative, but both twins are not equivalent (the ship experiences additional acceleration to changes the direction of travel).

So "… during the U-turn [of the twin-traveler's way] the plane of simultaneity jumps from blue [firm line here] to red [dotted line here] and very quickly sweeps over a large segment of the world line of the Earth-based twin".

By this way the "paradox" becomes be resolved – since, in full consistence with the SRT, the twin-homebody is older then the traveler one, though both twins thought during the travel that other twin must be younger.

From the consideration above follows, that only a rotation (180 degree here) of the traveler ship's direction results in that on Earth (and in a half of all Matter, though) everything becomes be older, on Earth – be older on this "segment". If the traveler's Earth-planet path was, e.g., 10 light years, then on Earth such the segment's aging is (let the ship's speed is near the speed of light) roughly 20 years. However from this "resolution" follows, also, that if the traveler will decide to return to the planet (near the planet, but when the ship's speed already is near speed of light), then after corresponding ∩-turn the traveler makes all on Earth younger again, including – he raise from the dead some people and give birth back to some others. But such events are evidently inconsistent, for example, with thermodynamics.

Another popular "resolution" of the paradox is as "the paradox arises since the traveler's and the homebody's frames aren't equivalent since the traveler's one was accelerated, and so is "non-completely inertial" frame". The flaw is evident – when the traveler returns to the homebody, he occurs simultaneously in both - homebody's and own – frames; so one should think, for example, also, that both twins' frames became be inertial and "non- inertial", since the traveler was accelerated at least four times; what is again illogical.

The VFL- theory is not contradictory, including it has not the twin paradox (the "paradox" evidently disappears if some preferred frame is postulated in a theory) – but had the uncertainty "what does this mathematical trick mean?" It seems that because this uncertainty the SRT was established (and is till now), in spite of the contradictories, as standard physical theory.

The informational model answers on the question above. Indeed, "relativistic effects" – "time dilation", "length contraction", etc., are, in *certain sense*, dynamical effects. They appear when some material object is impacted with changing of its momentum in generic reference frame; and, eventually, in an absolute reference frame, which is at rest relating to absolute Matter's spacetime FLE lattice and where Matter's mass value is minimal. Since the object always moves in the 4D spacetime with constant speed only, the impact results in the change of the object's direction and in a rotation of the object in the $(\tau, X)$ plain. Correspondingly its speed in the temporal direction becomes slower, internal processes in object slow down also. So there isn't a "time dilation", in reality, when measuring time by a clock there is a slowing down of the clock's rhythm. There isn't a "space contraction", in reality, when measuring of a moving rod length there is a result of measurement of the rod's projection on *X*-axis. As well as there aren't any "transformations" of spacetime since neither space nor time can be impacted. .

"Relativistic effects", in turn, are in accordance with the theory *only for rigid material bodies* and all known experimental tests of the SRT (in reality - the VFL- theory was tested) were successful till now only since systems "instruments + Earth", including "satellites +Earth" are rigid systems of scales and clocks. But if components of an experimental technique don't constitute rigid body, then the differences between the SRT and the VFLT become be essential, for example – it would be possible to measure the orbital speed of a couple of clocks in an Earth orbit by using data on distance between clocks and clocks' showing data only (more – see Shevchenko and Tokarevsky 2011, 2012), what is impossible if the SRT is true.

Besides note, however, that the success of the SRT was also due to the success of the GRT, where the postulate on the unity of space and time is very important, but it seems as rather possible that further development of the informational model will let to account for the gravity force correctly, but conserving at that the space/ time mutual independence.

*2.2.3. More about the material objects in the spacetime*

From above it is possible to suggest rather reasonable model (more see Shevchenko and Tokarevsky 2012), which allows correctly describe motion and interaction of material particles/ bodes; at that – which doesn't contain self-contradictions.

An impact in τ-direction on a FLE, which is flipping in this direction, creates "usual material" particle ("T-particle"), e.g., - an electron, a proton, etc. T- particles are inertial (have a non-zero mass) for impacts in spatial directions, and have zero rest mass in τ-direction, moving (if are at rest in absolute space) along τ-axis with speed of light. An impact in a spatial (*X*-) direction on a FLE, which is flipping in this direction, creates a "X-particle", e.g., - a photon that have zero rest mass in *X*-direction, moving in space with speed of light. T-particle can move along some lines in both - space and the time directions (in spacetime direction), when X-particle can move only in a space direction, with the FLE flipping point moving back and forth in τ- time direction; and – if a photon has "correct" spin in the space (=$\hbar$), the majority of T- particles have non –integer "space" spins – though it is rather probable that in the time direction a T- particle's angular momentum is "correct", i.e. is equal to $\hbar$.

So a T-particle at rest in space has the momentum

$$\vec{p}_0 = m_0 c \vec{i}_\tau .\quad(1)$$

After some spatial impact, when the particle obtained a spatial momentum, $\vec{p}_x$, its total momentum is

$$\vec{p}_{tot} = \vec{p}_0 + \vec{p}_x ;\quad(2)$$

and again it is equal $p_{tot} = mc$. It is easy to show that in Euclidian spacetime, where τ-axis is orthogonal to any space direction,

$$m = m_0 \frac{1}{(1-\beta^2)^{1/2}} \equiv m_0 \gamma ,\quad(3)$$

where γ is Lorentz factor. If the particle's spatial speed is $V$, spatial momentum is $p_x = mV = \gamma m_0 V$.

When a rigid body – let a rod - is at rest, its length is orthogonal to $\vec{p}_0$ (to the τ-axis).

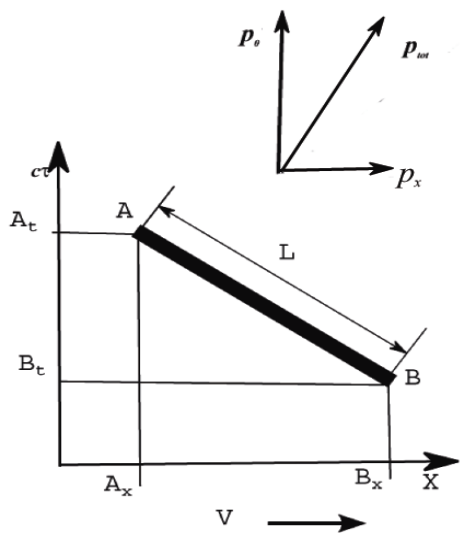

Fig. 2. A rod having the length *L* moves along *X*-axes with the speed *V*.

When the rod moves, after a spatial impact, in space, it remains be orthogonal to $\vec{p}_{tot}$ (Fig. 2); from this evidently follow Lorentz transformations (the rod moves in space from left to right, the rod's back end was in point *x*=0 in the generic inertial reference frame, *K*, at the time moment in K, $\tau = \tau' = 0$):

$$x = \gamma(x' + V\tau'), \qquad (4)$$

and

$$\tau = \gamma\left(\tau' + \frac{V}{c^2}x'\right), \qquad (5)$$

but the meaning of variables $x'$ and $\tau'$ is quite different comparing with the VFL-theory, and, moreover, with the SRT. That aren't some "spacetime points itself", that are results of the measurement of the distances along the rod and corresponding clocks' showings, if some (synchronized before the rod's acceleration) clocks were allocated along the rod's length; the scales and clocks on the rod are identical to the scales and clocks in the frame *K*. Any "frames' transformations" aren't defined in spacetime outside the rod.

Here isn't a "mathematical trick" (a convention) as in the VFLT. As well as here are no "pseudo Euclidian spacetime rotations" as in the SRT, when the rotations at a changing of frames is critically inherent to the SRT, since only at rotations (translations are inessential in this case) the spacetime interval

$$s^2 = -(\Delta x)^2 - (\Delta y)^2 - (\Delta z)^2 + c^2(\Delta t)^2, \qquad (6)$$

remains be invariant.

But the spacetime is absolute and cannot be rotated by any way.

This model differs, also, from so-called "Euclidian relativity" (refs. above) approach. The developers of the ER obtain the ER spacetime by not too correct method from the SRT spacetime by using the equation for the spacetime interval Eq. (6); and so obtain correct result – the ER-spacetime is Euclidian, there is absolute time ("Supertime", Nawrot, 2007), etc. But further they use the SRT meaning of the reference frame and frames transformations: any reference frame has infinite axis, the transformations between some frames are reduced to some rotations. As that was shown above, such a meaning is incorrect – if the spacetime is absolute it is incorrect to think that the spacetime points can be changed by some way; or simultaneously there can exist more then one frame with different directions of [coordinate] temporal axis.

This situation in the ER seems as follows from incorrect meaning of Lorentz transformations. So consider it more specifically.

The Galilean transformation for coordinate $x$ and frames $K$ and $K'$ is

$$x = V\tau + x' \qquad (7)$$

If one substitute $\tau'$ from Eq. (5) into Eq.(4), then obtain some analogue of Eq.(7)

$$x = V\tau + x'(1-\beta^2)^{1/2} \qquad (8)$$

The equation (8) one can obtain by two ways, however. On one hand it is true if the frame $K'$ is rotated relating to the frame K so, that the $x'$ axis is directed along the rod's axis. Further the measurement of the rod's length by a scale that had at rest, say, the length 1m, and at the measurement the scale constituted rigid system with the rod, results in that, e.g., the rod length is equal to $L$. Substitution of so obtained $x'$ coordinates to Eq. (8) gives true value of corresponding $x$ coordinate. On another hand one can measure the rod's length by "free" scale. In this case the scale measures the rod's projection on $X$-axes in $K$. But, since free scale turns out to be rotated in $(\tau, X)$ plain also, the measurement result will be the same, as in first case. The second measurement is correct – in contrast to the first one, because of in this case the $\tau$ -axes of both frames have the same direction, as that must be in the absolute spacetime. When at the first measurement it appears some temptation to set a rotated frame, what is made in existent "Euclidian relativity" papers, from what follow known problems in this approach.

## 3. Discussions and Conclusion

To conclude this consideration let us return to the Dowden's "list of problems" in Introduction (without semantic and neural problems, though). So:

• "What time actually *is?*" – The time is a realization of logical rule "Time" at ordering of causal sequences of events. Clocks in a moving rigid system measure clocks' proper [$\tau$-] time (cf. Dowden – time "is what clocks measure"). Clocks are always in some "coordinate time" points in the spacetime and clocks' showings are dilated relating to absolute time on the inverse Lorentz factor; which depends, in turn, on the speed of the clocks in spacetime (in an "absolute reference frame"). Since absolute speed in space of any material object isn't known till now, at experiments now only relative clocks' showings are measured.

• "Whether time exists when nothing is changing?" – Time, as the rule/ possibility that acts in whole Information Set, exists always. The [coordinate] time, as a *possibility* of some changes in Matter, exists always also, independently on Matter's existence. However, both rules *realize* themselves in existent Matter.

• "Whether there is a timeless substratum from which time emerges?"- There is no "a timeless substratum" [as well as no "a spaceless substratum"] from which time [space] can emerge; besides the logic, of course.

• "Why time has an arrow?" (and "How to analyze the metaphor of time's flow?") – Beginning resulted in appearance of Matter, when any Matter's object is an algorithm that always runs – there is "no friction" at FLEs' flipping, since FLEs are logically reversible (e.g., Margolus 2003). All existent material objects are some algorithms that are interruptedly running with constant operation rate what reveals them as always moving with constant speed (speed of light) in 4D spacetime. Even if some unchanging things were created at Beginning, they have come out from now observable Matter practically after first Matter's computer's [possibly Planck time] tact. So "time arrow (flow)" is a direction of sequence of logical cause-effect events – i.e., *absolute time* flow - that appear as result of Matter's computer's program non-interruptible code execution (cf. 't Hooft 2009:

> "… A correct description of all statistical features of the universe should be such that, right from the beginning, the probability distributions were best described by using the operators that we find useful today for describing the Standard Model. Those were eigenstates that always have been entangled…Thus, the entangled states are needed to describe our Universe from day one. *In any meaningful description of the statistical nature of our universe, one must assume it to have been in quantum entangled states from the very beginning.*" (G. 't Hooft's italics))

Though seems necessary to specify a little, what "Matter computer" is. In reality "primary" material objects, first of all – elementary particles, are closed algorithms that are [sufficiently to support Matter's stability] independent on External (for Matter) subsets in the Set "Information" and are capable to exist stably – if the algorithm started to run - on once own; Matter, generally speaking, is a [very] large set of little simple automata. But since everything in Matter is connected by universal gravity force, these automata become be "subroutines" that are united by gravity into some global program code. The problem "what is this code?" is outside the scope of this paper – but is rather interesting object for investigations in physics. As well as studies of Information Set as a whole become be main goal of the science, though.

• "Why time is one-dimensional and not two-dimensional?" The time rule in our Matter keeps the [absolute] time flow being equal for all material objects in Universe independently on anywhere and in what direction in spacetime any object moves. Thus on first sight it would be sufficient, for Matter evolution, to have one, i.e., absolute time, *t*; moreover this time seems as some "true time". But in reality the time is two-dimensional, since the coordinate time – or "the time that is what clock measure" exists also, and this time is evidently as something else then true (absolute) time, first of all – material objects move in absolute time in one direction, when in coordinate time they can move in both [±] directions. What is similar for both times – though absolute time isn't a coordinate - that there is the "absolute time point" on the coordinate [τ-] time axis, which uninterruptedly moves along this axis with the speed of light and so is equal to τ-time value of an object that is at rest in space. At that all Matter's objects are in the point *t simultaneously,* but are dispersed in whole spacetime, including along the τ- time axis. Such a situation happened, since for *given* Matter is necessary to have been possible at least two logical sequences in some cyclic algorithms - direct and reversal. And, though any logical sequence "cause-effect" has always one – "true time" direction, the reversal sequences are, in certain sense, "true" also; such sequences are executed in Matter as "coordinate time motion". So negative τ-time motion of a particle means that its code is reverse relating to corresponding particle; both particles constitute pair "particle and antiparticle".

Note, that in the SRT (in certain sense) the time is "two-dimensional" also, but here an analogue of the "absolute time", $t$, is the temporal coordinate, when an analogue of the coordinate time, $τ$, is a parameter of a body motion, "the proper time". Just this combination has required introducing so called "pseudo-Euclidian spacetime" in the theory.

Here a problem arises – why Matter's spacetime is so complex? An answer – or at least a part of the answer – follows from the fact that only in reversible algorithms transformations of information pass without energy losses (Margolus, 2003). So for whole Matter's evolution was enough to obtain some portion of energy only at Beginning. Why our Matter was made by such a way? – This problem is, of course, outside the scope of this paper.

- "Whether future time will be infinite?"
- "Whether there was time before the Big Bang event?"
- "Whether the future and past are as real as the present?"

This set of problems goes out the physics and requires returning to the metaphysical properties of the Information Set. So, in the Set "absolutely everything have happened already absolutely long time ago"; including total evolution of our Universe and Its Matter. The film "Matter in given Universe" was made with Planck [absolute] time step "absolutely long time ago" also and goes on "forever"; anybody now is only a player and observer in current film's version, when all film's pictures existed, exists and will exist (changing simultaneously for all Matter's objects) forever also. Thus *in certain sense*, e.g., the future and past are as real as the present; time was always before the [current] Beginning event, future time will be infinite. But in *current* film's version the past, present and future states of Matter are distinct, since the spacetime structure is discrete, when all Matter's objects are simultaneously in the same absolute time point, more correctly – in one time interval that is equal, rather possibly, to Planck time. At that the objects – as well as the objects' parts, since any object isn't point-like – can be in very different points in the coordinate time.

Note, also, that those "metaphysical properties" above reveal themselves somehow, as it seems, in existent mathematical models of spacetime, though the models differ from presented here one. A few examples: **[**Minkowski, cited in (Petkov 2002)]: "…We should then have in the world no longer *space*, but an infinite number of spaces [more correct – number of 4D spacetimes], analogously as there are in three-dimensional space an infinite number of planes." C. Gödel (cited in Zinkernagel 2011): "… reality consists of an infinity of layers of 'now' which come into existence successively" – these inferences rather possibly are some consequences, which (possibly implicitly for authors) ensue from the Set's structure.

The last considered here Dowden's problem is "What kinds of time travel are possible?" This problem doesn't relate only to science fiction stories; it was considered by Gödel and he obtained well known result that some spacetime + Matter configurations allow in general relativity theory an existence of time-like closed curves for material points. However, as it was pointed above, used in standard relativity theories spacetime model doesn't account for the "absolute time" correctly, when this time synchronizes all time moments in Matter and flow only in one direction by definition. So [at lest very probably] time-like curves – in spacetime + Matter system for current/ existent "film" - cannot be closed, and the answer about possibility of time travels is "No".

Finally some remark relating to the space. The situation in this case is similar to the case of the [coordinate] time. The space is a realization of logical rule "Space" at separating of fixed informational patterns. As well as the space, as the *rule* and a *possibility* of separating fixed informational patterns in Matter, is absolute and exists always.

As in time there is no "time dilation"; when in reality specific material objects, i.e., clocks, change their showing rate in accordance with changing of their speed along coordinate time direction after some external impact, there is no "space contraction". The observed/ measured "contraction" relates to material objects only, because of only material objects – in contrast to space or to time – can be impacted and so their parameters in spacetime can be changed. Sizes of a moving rigid system always are (if, of course the system isn't cracked at an impact when accelerating) the same at inertial motion as if the system is in the rest relating to the absolute / generic reference frame. But measured values of sizes of a moving rigid system, if are measured in generic (or other, where the system isn't at rest) become be "contracted". That happens because of (let the system is a rod having the length, $L$, and moving along $X$-axis), after an impact along a rod when the rod gets the speed $V$, the rod becomes be rotated in the plain ($\tau$, $X$). At that the rod's front end becomes be in the coordinate time's past, relating to the back end, on the Voigt-Lorentz decrement $\frac{-VL}{c^2}$, so the rod's projection on the $X$-axis of generic frame is equal to $L(1-\beta^2)^{1/2}$. Further – since any physical interaction, including any measurement *occurs only in space and in absolute time* (so independently on the coordinate time), the measurement of the moving rod's length results in "FitzGerald-Lorentz contraction".

For the inferences here it is critically important, that (i) -every material body moves with identical speed [of light] in the [absolute] 4D spacetime, and (ii) - the spacetime is Euclidian and the temporal axes is orthogonal to the spatial ones. Corresponding premise here is based on another premise – that FLE flips in any direction are logically independent. In the SRT four-velocity is identical for all material bodies in the Minkowski space (and is equal to speed of light) also, but here is no orthogonality of axes– on the temporal axis the *t*-component of the particle's momentum is placed, which is equal to the total momentum Eq. (2). Which, in reality, can have any direction in the 4D spacetime, including, for example, it can have negative τ-time component – all antiparticles move in negative τ-time directions relatively to positive (by convention) τ-time directions for particles.

But now known experimental data allow to make a reversal inference – from measured dependence of mass on the speed, which turns out to be in accordance with Eq. (3), follow both initial premises about the spacetime above, first of all – the orthogonality of the τ--axis to any stright line in the 3D space.

*Acknowledgements*

Authors are very grateful to Professor M. S. Brodin for support and useful discussions of the problems that were considered in this paper